\documentclass[a4paper,11pt]{article}
\pdfoutput=1 

\usepackage{jcappub} 

\usepackage[T1]{fontenc} 

\title{\boldmath General modified gravity with 21cm intensity mapping: simulations and forecast}

\author[a,b]{C. Heneka} 
\author[b]{and L. Amendola}

\affiliation[a]{Scuola Normale Superiore, Piazza dei Cavalieri 7, 56126 Pisa, Italy}
\affiliation[b]{Institut f\"ur Theoretische Physik, Ruprecht-Karls-Universit\"at Heidelberg, \\ Philosophenweg
16, 69120 Heidelberg, Germany}

\emailAdd{caroline.heneka@sns.it}
\emailAdd{amendola@thphys.uni-heidelberg.de}

\abstract{Line intensity mapping opens up a new and exciting window for probing cosmology and fundamental physics during the Epoch of Reionisation, extending to redshifts previously untested by galaxy surveys. The power spectra of these line fluctuations are a promising tool to test gravity over a large range of scales and redshifts.
We simulate cosmological volumes of 21cm fluctuations in general parametrisations of modified gravity, in order to calculate the corresponding power spectra, where additional parameters are the initial condition of matter perturbations $\alpha$ and the scale-dependent modified gravity parameter $Y$ (also known as $G_\mathrm{eff}$) that measures deviations from GR in the Poisson equation. We show the impact of these model-independent modifications of gravity, to either delay or expedite reionisation. For the 21cm intensity mapping survey to be performed by the SKA mission, we forecast the ability of line intensity mapping to constrain the parameters $Y$ and $\alpha$ at redshifts $z=6-11$, where $Y$ is assumed constant during this epoch (but without requiring constancy at all times). In our most conservative scenario, the $Y$ parameter can be constrained at the tens of percent level, while for improved modelling of foregrounds as well as of the (mildly) non-linear regime, up to sub-percent level constraints are attainable. We show the impact of jointly estimating reionisation model parameters and corresponding parameter correlations, as well as of foreground removal. We note, that tomography is crucial to break degeneracies and for constraints not to degrade significantly when adding reionisation model parameters, with most constraining power coming from the redshift bins $z=7-10$ where the shape of the 21cm power spectrum is evolving fastest.
}

\keywords{cosmological parameters from LSS, modified gravity, cosmological simulations, dark energy theory}

\arxivnumber{1805.03629}

\begin{document}
\maketitle
\flushbottom

\section{Introduction}
\label{sec:intro}
The Cosmic Microwave Background (CMB) has been and still is a powerful probe of cosmology~\cite{2016A&A...594A..14P} --- but soon intensity mapping, a probe similar conceptually, where intensity fluctuations of line emission over a large range of redshifts and scales are mapped out, will complement more traditional probes. Intensity mapping will go beyond the measurements at low redshifts of single astrophysical objects to constrain cosmology, as for example surveys of galaxies and clusters of galaxies, that successfully constrained cosmology for both the standard scenario of $\Lambda$CDM and extensions like modified gravity at lower redshift~\citep[for example][]{Rapetti12, 2015MNRAS.446.2205M,Taddei:2016iku}. Also, measurements of intensity fluctuations will enable us to push to redshifts as high as the Epoch of Reionisation (EoR)~\citep{2017ApJ...848...52H}. 

At the EoR the radiation of the first stars and galaxies ionises again the cold and neutral Intergalactic Medium (IGM) around them, that is mostly made up of hydrogen. Starting around a couple of million years after the Big Bang, reionisation currently is constrained to be completed around $z\sim 6$, when the medium is fully reionised, by observations of the Ly$\mathrm{\alpha}$ forest toward quasars~\citep{2006AJ....132..117F, 2015MNRAS.447..499M}. Both the reionisation model, regarding for example ionising sources, spatial structure and its onset, and cosmology, regarding the relative importance of the energy components in our Universe that governs structure formation, are very uncertain or even unconstrained during this epoch.

Upcoming mission like the Square Kilometre Array (SKA) will soon provide first constraints on both astrophysics and cosmology by detecting power spectra of 21cm fluctuations at redshifts of reionisation, mapping most of the sky~\citep{2015aska.confE...1K}.\footnote{https://skatelescope.org/} Here the so-called 21cm line is the forbidden spin-flip transition of neutral hydrogen, whose fluctuations will be crucial in following structure formation and the expansion of ionised regions during reionisation, due to hydrogen being extremely abundant and the 21cm line tracing the neutral IGM.
Also, interesting synergies with other upcoming missions like Euclid~\citep{Amendola:2012ys} are to be expected in terms of constraining cosmology over cosmic times.

These probes of new redshifts and scales are crucial when trying to explain cosmic acceleration, first discovered with measurements of SN Ia luminosity distances~\citep{Riess:1998cb,Perlmutter:1998np} and since then consistently confirmed. Besides the standard picture of Cold Dark Matter (CDM) plus a cosmological constant $\Lambda$ to fit observations, additional fundamental fields can be added to General Relativity (GR), in order to explain cosmic acceleration. These change the effective gravitational potential and thus modify gravity. The hunt for these modifications has been a question central to cosmology. A wealth of models has been proposed (see e.g.~\citep{amendola_tsujikawa_2010,Koyama:2015vza}), and it is advisable to search for signatures of their modifications to gravity in a manner as model-independent as possible.

For many models beyond $\Lambda$CDM, modifications at the background level can be effectively parametrised as a fluid with an equation of state, $w$, which in general is time-dependent. At linear perturbation level and for sub-horizon scales, scalar modifications  to $\Lambda$CDM can be parametrised by two time- and space-dependent functions. These are the modification to the standard Poisson equation $Y\left( a,k\right)$, with scale factor $a$ our time variable and wavenumber $k$ denoting our space dependence in Fourier space, which in the scale-independent case often is also called the effective gravitational constant $G_\mathrm{eff}$, or effective gravitational coupling to non-relativistic matter. The second function is the anisotropic stress $\eta\left( a,k \right)$. It is the ratio between the Newtonian gravitational potentials $\Phi$ and $\Psi$ that are introduced as scalar degrees of freedom in the perturbed Friedmann-Lemaitre-Robertson-Walker (FLRW) metric. Deviations from the standard GR case of $Y=\eta=1$ are an indicator of modifications to gravity, where deviations from unity for $Y$ can also hint to clustering dark energy (as measurable e.g. by galaxy clusters~\cite{Heneka:2017ffk}), while lensing is sensitive to deviations from unity for differences in the Newtonian potentials as parametrised by $\eta$. Especially for the so-called Horndeski class of scalar-tensor models~\citep{1974IJTP...10..363H}, the form of the effective gravitational coupling to matter $Y$ takes on a specific form in the quasi-static limit, whose characteristic scale-dependence is a  signature for modifications to gravity of the Horndeski type. Scale- and time-dependence of $Y$ have not been constrained well so far (constraints e.g. from galaxy redshift surveys are expected to be weak~\cite{Taddei:2014wqa}). Examples of a non-constant functional form were explored in~\cite{Nesseris:2017vor} for growth rate data and in~\cite{2016A&A...594A..13P} for CMB data. But e.g.~\cite{Brax:2012cr,Lima:2016npg} already found 21cm intensity mapping during reionisation to be a powerful discriminator of modified gravity models, like~\cite{2013PhRvD..87f4026H,Bull:2015lja} found at lower redshift.

Besides an effective strength for gravity parametrised with $Y$, also our lack of knowledge of the initial conditions for matter perturbations can be parametrised. In fact, in modified gravity models, the initial conditions at high redshift are not necessarily those of a pure CDM Universe where initial conditions are given by $\delta'_{in}=\delta_{in}$. To parametrise for this possibility, we introduce the parameter $\alpha$, that is unity in a pure matter dominated early Universe, but might deviate or even be scale-dependent in modified cosmologies. Examples are Brans-Dicke models where $\delta'_{in}$ depends on the coupling $\omega$~\cite{doi:10.1143/PTP.42.544} and coupled dark energy with non-negligible dark energy densities at early times~\cite{Amendola:2000uh}.

When modelling the EoR, for the redshifts and scales of interest, it is mostly a good enough approximation to work at linear level in perturbations, making this an interesting playground to test modifications to gravity manifesting themselves in $Y$ and $\alpha$.
A lot of work has gone into modelling and preparing detections of the 21cm signal and power spectrum of fluctuations, using semi-numerical simulations, such as 21cmFAST~\citep{Mesinger10}. These simulations enable a full and consistent parameter exploration, as compared e.g. to hybrid N-body approaches with hydrogen added later to the simulation as in~\cite{Carucci:2017cnn}.
We here expand these simulations to the general modified gravity case, evolving the linear matter growth in models with both $Y$ and $\alpha$, opening a unique avenue to test the expected 21cm signal, therefore testing the underlying growth of structures, at the redshifts of reionisation.

The goal of this paper is to extend standard reionisation simulations set in a $\Lambda$CDM cosmology with GR to general modifications of gravity and explore what 21cm signal we can attain to trace the growth of structures in these models, giving a unique handle to constrain high-redshift deviations from GR. We finish by forecasting constraints on $Y$ and $\alpha$ attainable with 21cm intensity mapping during the EoR.

\section{Simulating 21cm fluctuations in general modified gravity}

\subsection{Background and growth}\label{sec:growth}

The background expansion is given by the dimensionless Hubble parameter $E\left(a\right)=H\left(a\right)/H_0$, with Hubble parameter $H\left(a\right)$ and Hubble constant $H_0$, for present-day matter density $\Omega_\mathrm{m,0}$, radiation density $\Omega_\mathrm{r,0}$ and time-evolving dark energy equation of state $w\left( a\right)$ as a function of the scale factor $a$, as
\begin{equation}
E^2\left( a\right)=\Omega_\mathrm{m,0}a^{-3} + \Omega_\mathrm{r,0}a^{-4} + \left( 1-\Omega_\mathrm{m,0}\right)e^{3\int\left( 1+w\left( a'\right)\right)/a' da'} .
\end{equation}
We parametrise the time-evolution of $w$ as a Chevallier-Polarski-Linder form with $w\left(a\right)=w_0 + w_\mathrm{a} \left( 1-a\right)$~\citep{Chevallier:2000qy}. The background then can be written as
\begin{equation}
E^2\left( a\right)=\Omega_\mathrm{m,0}a^{-3} + \Omega_\mathrm{r,0}a^{-4} + \left( 1-\Omega_\mathrm{m,0}\right)a^{-3\left( 1+w_0+w_\mathrm{a}\right)}e^{-3 w_\mathrm{a}\left( 1-a\right)} .
\end{equation}
The parameter $Y$, the effective gravitational strength, enters at the linear perturbation level when calculating the growth of matter perturbations.
The linear growth of perturbations for a general modification of gravity follows the equation~\citep{Amendola:2012ky}
\begin{equation}
\delta''_\mathrm{m} + \left( 2+\frac{E'}{E}\right)\delta'_\mathrm{m} = \frac{3}{2}\frac{\delta_\mathrm{m}}{a^3E^2}\Omega_\mathrm{m}Y ,
\label{eq:pert2}
\end{equation}
with $Y=1$ for $\Lambda$CDM. In more general scenarios the function $Y=Y\left(a,k\right)$ can vary with time and scale. The prime denotes derivatives after $\log a$. Fiducial model parameters, if not stated differently, are $\sigma_8=0.815$, $h=0.678$, $\Omega_\mathrm{r}=8.6\times10^{-5}$, $\Omega_\mathrm{m,0}=0.308$, $w_\mathrm{0}=-1$, $w_\mathrm{a}=0$, $Y=1$ and $\alpha =1$.

The initial condition parameter $\alpha=\delta'_{in}/\delta_{in}$ is equal to unity for early matter domination, but is as well a free parameter in more general scenarios that allow for deviations from early matter domination through modifications of gravity or the presence of early dark energy. In principle, the parameter $\alpha$ can also be a scale-dependent function. We will treat $\alpha$ in the following as a free constant parameter when deriving constraints on modifications of gravity. 

As it represents the deviation from the GR case in the Poisson equation for non-relativistic species, i.e., the effective gravitational constant for matter, the parameter $Y$ can be expressed as the function
\begin{equation}
Y\left( z,k \right) = -\frac{2k^2\Psi}{3\left( aH \right)^2 \Omega_\mathrm{m}\delta_\mathrm{m}}  .
\end{equation}
For the Horndeski type of scalar-tensor models, in the quasi-static limit where we neglect time-variations at the perturbative level, $Y$ can be expressed as~\citep{Amendola:2012ky}
\begin{equation}
Y = h_1 \frac{1+\left(k/k_p \right)^2 h_5}{1+\left(k/k_p \right)^2 h_3} ,
\end{equation}
with time-dependent functions $h_1$, $h_3$, $h_5$, and (arbitrary) pivot scale $k_p$. We will in the following investigate the case $Y$ constant during the EoR (but which is not required to be constant at all times), to detect deviations from $Y=1$ of whatever form possible. 

We evolve linear perturbations by making us of the growth function defined as $G\left( a,k\right)=\delta_m\left( a,k\right)/\delta_m\left(1,k\right)$ and calculated via eq.~(\ref{eq:pert2}). For example the rms variation of the matter density field within 8~$h^{-1}$Mpc spheres, $\sigma_8$, at arbitrary $a$ is then given as $\sigma_8\left( a\right)=G\left( a\right)\sigma_8$. We cross-checked the growth function normalised to present time, both evolved via eq.~(\ref{eq:pert2}) and by using a $\Lambda$CDM fitting function~\cite{Liddle:1995pd}, that has previously been used in 21cmFAST\cite{Mesinger10}, for parameter values $w=-1$ and $\Omega_\mathrm{m,0}$ varying. Fitting function and the cosmology-dependent growth evolution newly implemented via eq.~(\ref{eq:pert2}) in 21cmFAST agree well. 

In the following,  in order to comply with CMB observations at high redshift, we chose for overall normalisation to normalise the growth at the scale factor or redshift of the CMB. All models then exhibit the same growth, or alternatively $\sigma_8$, at the time of recombination with $z_{cmb}\approx1090$, with the standard $\Lambda$CDM scenario as our reference model. We thus obtain for different modified gravity parameters a different $\sigma_8$ at present time, with our reference value being $\sigma_8=0.815$ in the $\Lambda$CDM case in accordance with Planck measurements~\citep{2016A&A...594A..13P}.

\subsection{Simulations}\label{sec:sim}
Here we show simulations of cosmological volumes of 21cm line emission, that can be targeted by intensity mapping experiments at relevant redshifts during the EoR. Due to the abundance of hydrogen and the property of the 21cm signal to trace the neutral medium, which tends to be underdense and reionised last, intensity mapping of the 21cm signal is a promising tool to follow the formation of structures during reionisation.

Maps of 21cm line emission and the corresponding ionisation fields can be efficiently simulated with seminumerical codes, while at the same time showing good agreement to {\it N}-body codes coupled with radiative transfer as well as analytical modelling at the redshifts of reionisation.
We use the seminumerical code 21cmFAST,\footnote{https://github.com/andreimesinger/21cmFAST} modified to incorporate growth as evolved in our general modified gravity scenario, to create linear density, linear velocity and evolved velocity fields at first order in Langrangian perturbation theory~\citep[Zel'dovich approximation,][]{Zel70}, as well as ionisation fields in the framework of an excursion set approach. 
In this approach overdensities are filtered at different radii, where those regions are assigned to be ionised whose overdensity is higher than a collapsed fraction set by an ionising efficiency $\zeta$. This parameter $\zeta$ is an effective reionisation model parameter which we choose in accordance with current bounds. In principle it depends on a combination of underlying possible reionisation model parameters as e.g. the escape fraction of UV radiation of galaxies. 

We then can calculate the 21cm brightness offset temperature $\delta T_\mathrm{b}$ between spin gas temperature $T_\mathrm{S}$ and CMB background temperature $T_{\gamma}$, where the spin temperature is measured by the ratio of the occupation of Boltzmann levels for the forbidden spin flip transition of neutral hydrogen in its ground state.
The 21cm brightness temperature offset $\delta T_\mathrm{b}$ at redshift $z$ and position $\bf{x}$ is obtained via
\begin{align}
\delta T_\mathrm{b} \left({\bf x}, z \right) &= \frac{T_\mathrm{S}-T_{\gamma}}{1+z}\left(1- e^{-\tau_{\nu_0}} \right) \nonumber \\
& \approx  27x_\mathrm{HI}\left( 1+\delta_\mathrm{nl}\right)\left( \frac{H}{\mathrm{d}v_\mathrm{r}/\mathrm{d}r + H}\right)\left( 1- \frac{T_{\gamma}}{T_\mathrm{S}}\right) \nonumber \\ & \vspace{0.2cm}\times \left( \frac{1+z}{10} \frac{0.15}{\Omega_\mathrm{m,0} h^2}\right)^{1/2} \left( \frac{\Omega_\mathrm{b} h^2}{0.023}\right) \mathrm{mK} , \label{eq:Tb}
\end{align}
where redshift $z$ is related to observed frequency $\nu$ as $z = \nu_0/\nu -1$, with optical depth $\tau_{\nu_0}$ at rest frame frequency $\nu_0$, ionisation fraction $x_\mathrm{HI}$, non-linear density contrast $\delta_\mathrm{nl}=\rho / \bar{\rho}_0 -1$, Hubble parameter $H\left( z\right)$, comoving gradient of line of sight velocity $\mathrm{d}v_\mathrm{r}/\mathrm{d}r$, as well as present-day matter density $\Omega_\mathrm{m,0}$, present-day baryonic density $\Omega_\mathrm{b}$, and Hubble factor $h$. We use this relation eq.~(\ref{eq:Tb}) for the 21cm brightness temperature in the post-heating limit $T_{\gamma} \ll T_\mathrm{S}$ when the CMB temperature is much lower than the spin gas temperature. During the heating epoch that precedes reionisation in principle the full evolution of the spin gas temperature needs to be taken into account.

Temperature fluctuations $\delta_{21} \left({\bf x},z\right)$ on the simulated grid at position $\bf{x}$ and redshift slice $z$ are calculated as
\begin{equation}
\delta_{21} \left({\bf x},z\right) = \frac{\delta T_\mathrm{b}\left( {\bf x},z\right)}{\bar{T}_{21}\left( z\right)}-1
\end{equation}
with average 21cm brightness temperature $\bar{T}_{21}\left( z\right)=<\delta T_\mathrm{b}>_{{\bf x}}$. 

In the following, the fiducial reionisation model parameters are chosen as
\begin{align}
R_\mathrm{mfp} &=20 \, \mathrm{Mpc}, \  T_\mathrm{vir}=3\times10^4\, \mathrm{K}, 
 \ \zeta=20 ,
\label{eq:iparam} 
\end{align}
where $R_\mathrm{mfp}^\mathrm{UV}$ is the mean free path of ionising radiation, $T_\mathrm{vir}$ is the typical halo virial temperature and $\zeta$ the ionising efficiency, in order to reproduce a realistic reionisation history; our model is with an optical depth of $\tau_\mathrm{fid} \sim 0.057$ in agreement with the optical depth of the CMB as measured by Planck~\citep{2016A&A...594A..13P}. We also choose the relevant parameters for the heating history, the efficiency of X-ray heating $\zeta_\mathrm{x}$ and the mean baryon fraction in stars $f_{*}$, as $\zeta_\mathrm{x} = 2\times10^{56}$ and $f_{*} =0.05$. 

\begin{figure}
\begin{center}
\includegraphics[width=0.32\columnwidth]{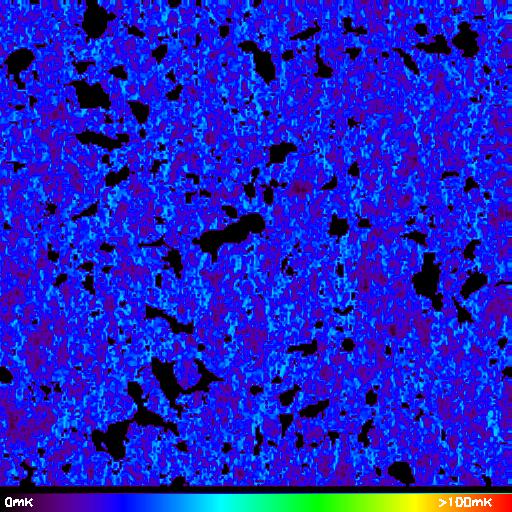}
\includegraphics[width=0.32\columnwidth]{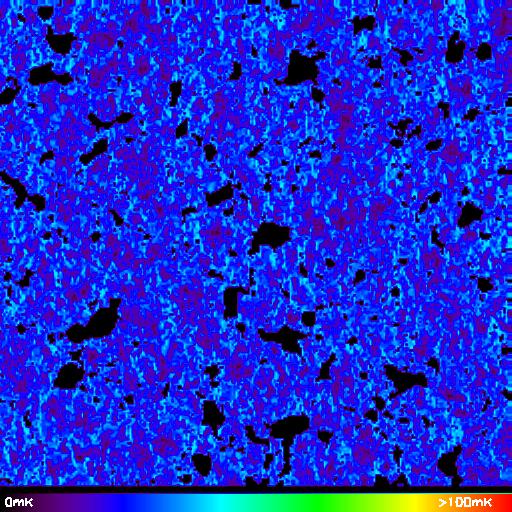}
\includegraphics[width=0.32\columnwidth]{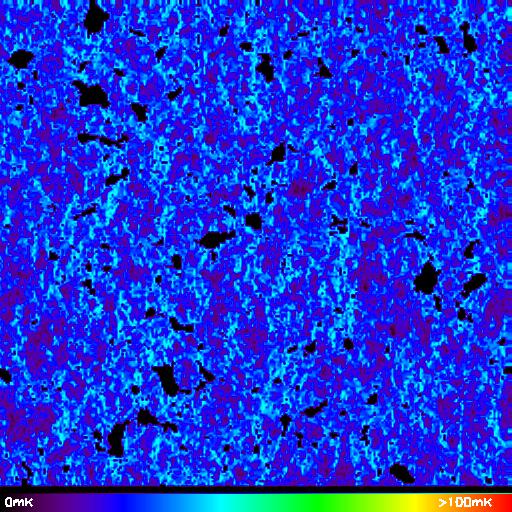}
\includegraphics[width=0.32\columnwidth]{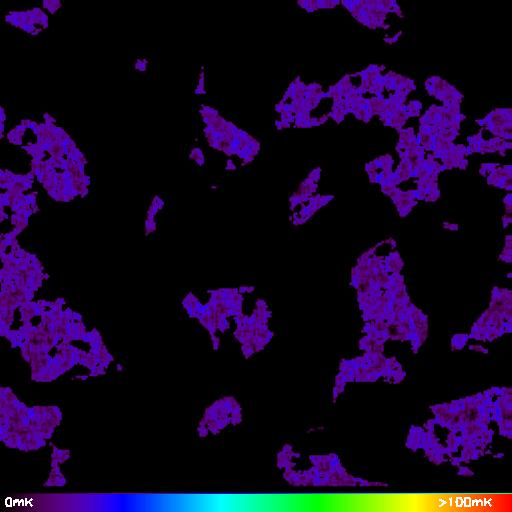}
\includegraphics[width=0.32\columnwidth]{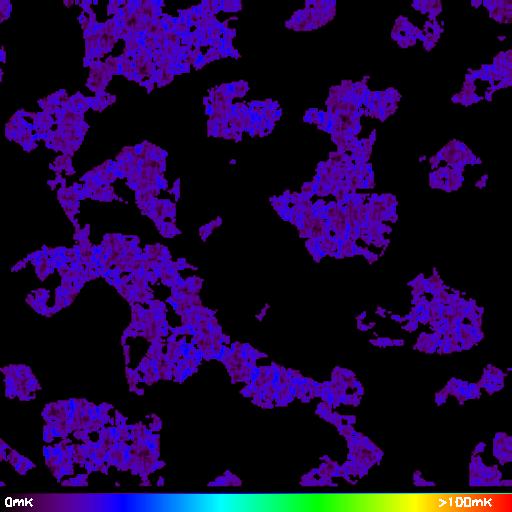}
\includegraphics[width=0.32\columnwidth]{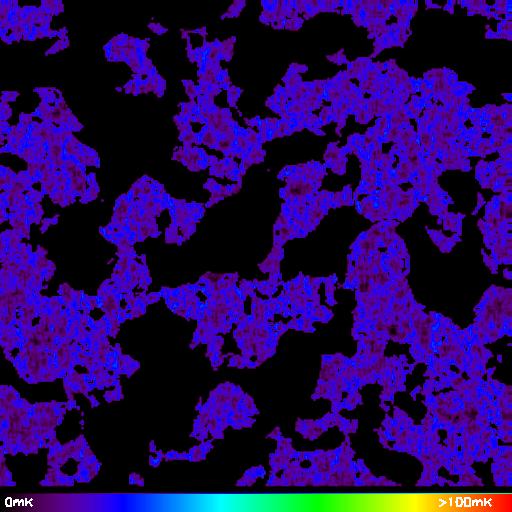}
\caption{Examples of 300$\,$Mpc simulation boxes of 21cm emission at redshift $z=10$ (top row of panels) and $z=7$ (bottom row of panels), for our fiducial cosmology with $\alpha=1$, but varying $Y=1.01$ (left), $Y=1.00$ (middle) and $Y=0.99$ (right); the growth function was normalised to the same value at $z_{CMB}$ for all models. The $Y$ parameter is chosen such that the corresponding values of $\sigma_8$ are $\sigma_8=0.785$ and $\sigma_8=0.846$ for $Y=0.99$ and $Y=1.01$, respectively, close to present-day observational bounds, with our fiducial value $\sigma_8=0.815$ for $Y=1.00$.}
\label{FIG:Box21_Y_CMB}
\end{center}
\end{figure}

In figure~\ref{FIG:Box21_Y_CMB} simulated boxes of 300$\,$Mpc box size for fluctuations in 21cm brightness temperature are depicted, sliced at redshift $z=10$ and $z=7$. Going from $z=10$ (top panels) to $z=7$ (bottom panels), i.e., from high to low redshift, the growth of ionised patches with negligible 21cm emission and therefore a lower signal in 21cm emission becomes obvious, as 21cm emission is tracing neutral hydrogen.  We can see going from left to right, that depending on the value of $Y$, reionisation has progressed more at redshift $z=10$ and $z=7$, respectively, with $81\%$ and $18\%$ of the medium still neutral for $Y=1.01$ (left panels) compared to the fiducial model of $Y=1.00$ (middle panels) with $87\%$ and $27\%$ neutral in hydrogen, or is less advanced for $Y=0.99$ (right panels), with a fraction of $91\%$ and $45\%$ of hydrogen still neutral. In order to be roughly consistent with present-day constraints on $\sigma_8$ we chose for display besides the fiducial case of $Y=1.00$ and $\alpha=1$ with $\sigma_8=0.815$ the parameter values $Y=0.99$ and $Y=1.01$ with $\sigma_8=0.785$ and $\sigma_8=0.846$, respectively. This exemplary deviation from GR as parametrised by $Y$ is still well within bounds from Big Bang nucleosynthesis (BBN), which allow for a deviation of 10--20$\%$~\cite{Copi:2003xd,Bambi:2005fi}.

The corresponding optical depths are $\tau \sim 0.052$ for $Y=0.99$ and $\tau \sim 0.062$ for $Y=1.01$, as compared to our $\tau_\mathrm{fid} \sim 0.057$ (with a larger $\tau$ implying a higher redshift of reionisation), still within $1\sigma$ bounds of Planck. The required precision to detect such a shift is forecasted to be reached already by the CMB Stage 3 CLASS experiment~\cite{2016SPIE.9914E..1KH}, and Stage 4 CMB experiments like the space missions PIXIE and LiteBIRD even aim at cosmic-variance limited measurements of $\tau$ with a precision of 0.002~\cite{2011JCAP...07..025K,Matsumura:2013aja}. Note though, that probes of the 21cm line provide a measurement of $\tau$ independent from the CMB, pushing limits further down and breaking CMB degeneracies~\cite{Liu:2015txa}. 

In addition, as an example of how MG affects the reionisation history, we show in appendix~\ref{sec:extreme} simulation boxes of 21cm emission for two more extreme values of the parameter $Y$ that affects the strength of gravity. These extreme scenarios are ruled out though by present-day constraints, for example, on the end of reionisation and CMB optical depths, as well as constraints on $\sigma_8$.

\subsection{21cm power spectra}\label{sec:Pk}
In this section we show examples of power spectra for simulated 21cm line emission at different strengths of the modified gravity parameter $Y$. The power spectra shown correspond to the simulation boxes depicted in the previous section. We also perform an error calculation for the upcoming SKA interferometer to estimate the detectability of these spectra and show that the effect of a modified Poisson equation and therefore strength of gravity can be targeted by intensity mapping at high redshifts during the EoR at high precision.

For fluctuations $\delta_{21}$ in 21cm brightness offset temperature, we define in the following the dimensionless 21cm power spectrum as 
$\tilde{\Delta}_{21}\left( k\right) = k^3/\left( 2\pi^2 V\right) \left< |\delta_{21}|^2\right>_k$ and the dimensional power spectrum as $\Delta_{21}\left( k\right) = \bar{T}_{21 }^2 \tilde{\Delta}_{21}\left( k\right)$, with mean 21cm brightness offset temperature $\bar{T}_{21 }$. As mentioned previously in section~\ref{sec:growth}, we normalise the growth to the time of recombination at $z_{cmb}\approx1090$, so that for different modified gravity parameters we obtain a different $\sigma_8$ at present time, with our reference value being $\sigma_8=0.815$ in the fiducial case of $Y=1.00$.

The error on the 21cm power spectrum is estimated assuming for instrument specifics a SKA stage 1 intensity mapping experiment (SKA-LOW, see table~\ref{tab:exp}), including cosmic variance, thermal noise and limited instrumental resolution.
Including these effects the variance for our 21cm power spectrum estimate for angle $\mu$ between the line of sight and mode $k$ can be expressed as~\citep{2006ApJ...653..815M}
\begin{equation}
\sigma^2_{21} \left( k,\mu\right) = \left[  P_{21}\left( k,\mu\right) + \frac{T_\mathrm{sys}^2 V_\mathrm{sur} \lambda_{21}^2}{B\,t_\mathrm{int}n\left( k_{\perp}\right) A_\mathrm{e}}W_{21}\left(k,\mu \right) \right] , \label{eq:sigma21}
\end{equation}
where the first term includes cosmic variance, the second term the thermal noise of the instrument, and the window function $W_{21} \left( k,\mu \right)$ takes into account the limited spectral and spatial instrumental resolution in parallel and perpendicular modes. For an SKA stage 1 type instrument, we take the instrument specifications as listed in table~\ref{tab:exp}, together with an effective survey volume of $V_\mathrm{sur}=\chi^2\Delta\chi\left( \lambda_{21}\left( z\right)^2/A_\mathrm{e}\right)$, for redshifted 21cm wavelength $\lambda_{21}\left(z \right)$ as well as comoving distance and survey depth $\chi$ and $\Delta \chi$. 
The total variance $\sigma^2\left(k \right)$ of the full spherically averaged power spectrum is the sum over all angles $\mu$, divided by the number of modes per bin; here we explicitly counted the number of modes for each bin. Note that our assumption of constant number density of baselines $n_{\perp}$ tends to underestimate the error bars for high $k$ values, as $n_{\perp}$ tends to decrease for longer baselines.

 \renewcommand{\arraystretch}{1.0}
\setlength{\tabcolsep}{18pt}

    \begin{table}
    \centering
    \begin{tabular}{| c| c| c| c| c| c|c |}
     \hline
 $\nu_\mathrm{res}$  &  $l_\mathrm{max}$ & $T_\mathrm{sys}$ & $t_\mathrm{int}$ & B (z=8) & $A_\mathrm{e}$ (z=8) & $n_{\perp}$  \\
    (kHz) & (cm) & (K) & (hrs) & (MHz) & (m$^2$) & \\
\hline 
 3.9 &  $10^5$  & 400 & 1000 & 8  & 925 & 0.8 \\ 
 \hline
         \end{tabular}
\caption{Instrument specifications for 21cm survey: SKA stage 1. See section \ref{sec:Pk} for details on error calculations; specifications taken from~\cite{2015PritchardSKA,Chang:2015era}. Given are the spectral resolution $\nu_\mathrm{res}$, the maximum baseline $l_\mathrm{max}$, the instrument system temperature of $T_\mathrm{sys}$, the total observing time time of $t_\mathrm{int}$, the survey bandwidth $B$, the effective area $A_\mathrm{e}$ and the average number density of baselines $n\left( k_{\perp}\right)$ for mode $k_{\perp}$ perpendicular to the line-of-sight.}
    \label{tab:exp}
    \end{table}

When indicated we remove as well the so-called 21cm foreground wedge~\citep{MoralesWedge12} where foregrounds and instrument systematics due to leakage in the 21cm radio window dominate the signal. This foreground wedge is defined for the cylindrically averaged 2D power spectrum as
\begin{equation}
k_{\parallel} \leq \frac{\chi\left( z\right) E\left(z\right) \theta_0}{d_\mathrm{H}\left( 1+z \right)} k_{\perp} \,, \label{eq:wedge}
\end{equation}
for parallel modes $k_{\parallel}$ and orthogonal modes $k_{\perp}$, and with characteristic angle $\theta_0$, comoving distance $\chi\left( z\right)$ and Hubble distance $d_\mathrm{H}$ to determine the slope of the wedge. We take the characteristic angle as $10$ degrees, corresponding to the assumption that contaminations from residual sources are mostly limited to the primary beam, or the instrument field of view.

\begin{figure}[ht!]
\begin{center}
\includegraphics[width=0.48\columnwidth]{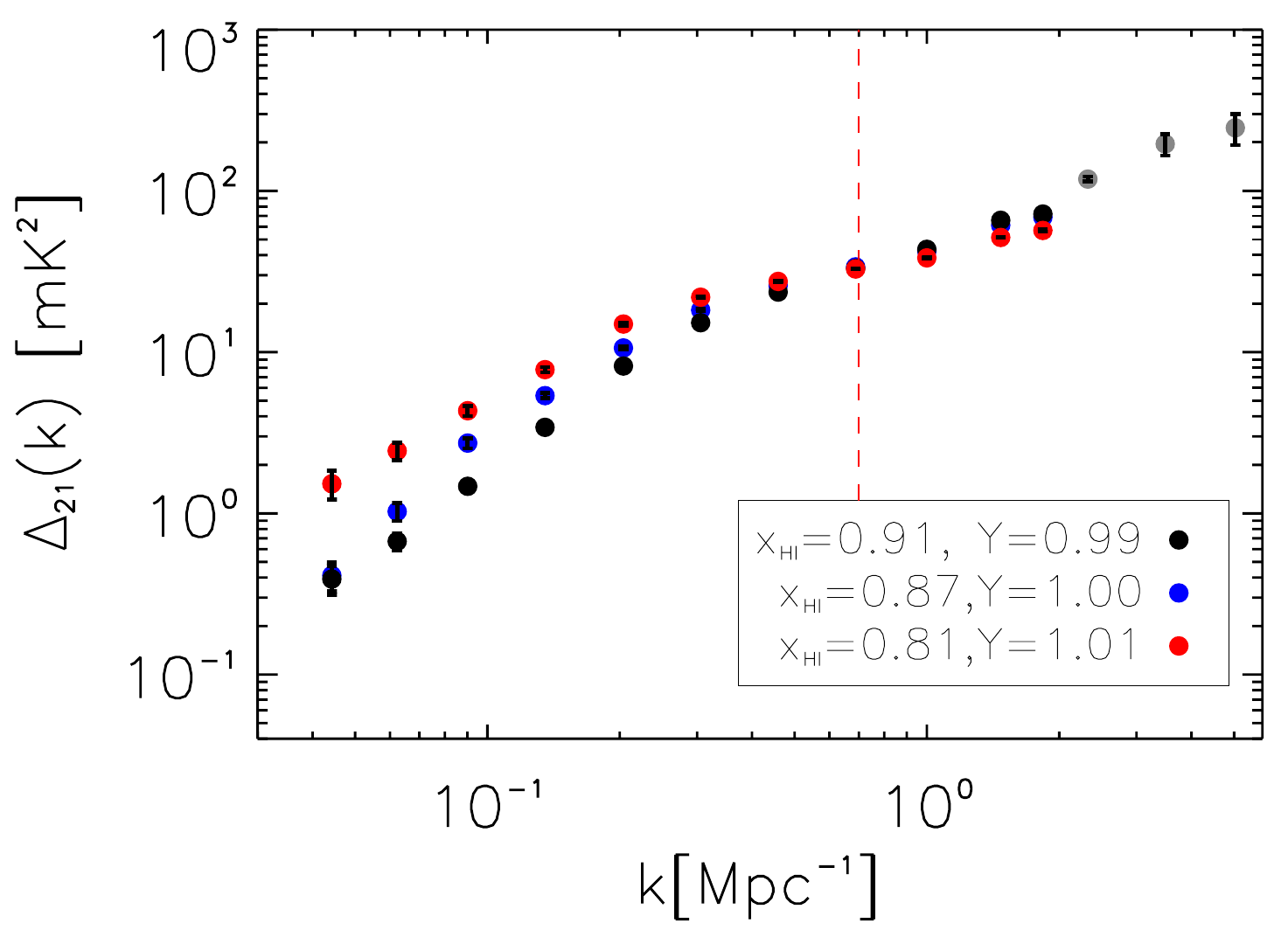}
\includegraphics[width=0.48\columnwidth]{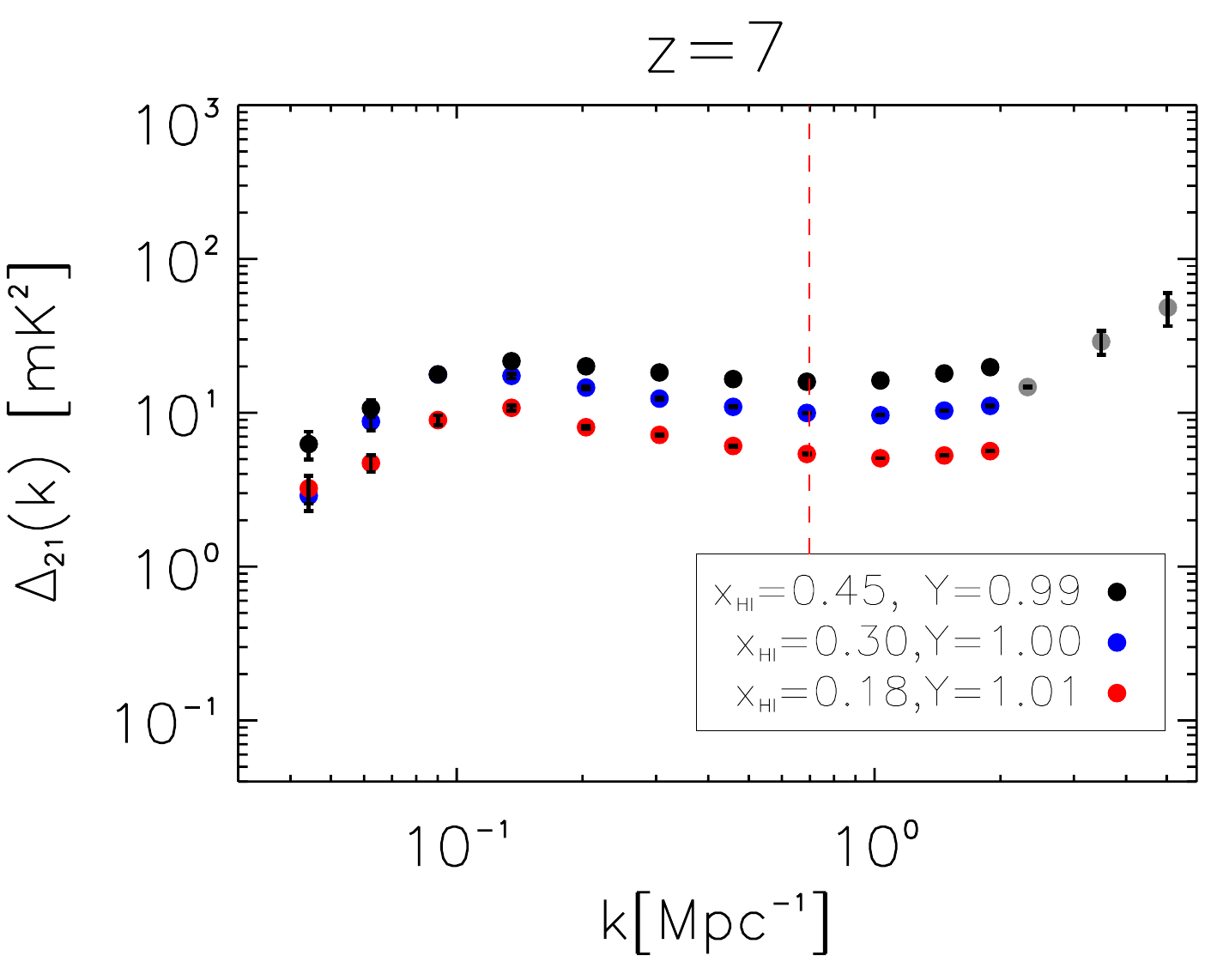}
\caption{Dimensional 21cm power spectrum at redshift $z=10$  and $z=7$ with errors bars calculated for a SKA stage 1 intensity mapping probe (see table~\ref{tab:exp}), including cosmic variance, thermal noise, limited instrumental resolution and foreground wedge removal, for parameter values $w_0=-1$, $w_\mathrm{a}=0$, $\Omega_\mathrm{m,0}=0.308$, $\alpha=1$ and varying $Y$ as indicated. $x_\mathrm{HI}$ denotes the mean neutral fraction; the growth function was normalised at $z_{CMB}$, where the $Y$ parameter is chosen such that the corresponding values of $\sigma_8$ lie with $\sigma_8=0.785$ and $\sigma_8=0.846$ for $Y=0.99$ and $Y=1.01$, respectively, close to present-day bounds, with our fiducial $\sigma_8=0.815$ for $Y=1.00$. The red vertical dashed lines depict roughly the onset of the non-linear regime at the corresponding redshifts, while the high-k grey points belong to a fiducial $Y=1.00$ higher-resolution simulation and depict the regime where limited frequency resolution and shot-noise become important.}
\label{FIG:P21_Y_CMB}
\end{center}
\end{figure}

Figure~\ref{FIG:P21_Y_CMB}  shows the 21cm power spectrum at $z=7$ and $z=10$ for values $Y=0.99$, $Y=1.01$ and the GR expectation $Y=1.00$, i.e., for different values of the modified gravity parameter $Y$. The power spectra are derived from the boxes of  simulated 21cm emission as shown in section~\ref{sec:sim}. The error bars are assigned according to our error estimate for survey specifications as described above and depicted in table~\ref{tab:exp}, including cosmic variance, thermal noise and limited instrumental resolution, as well as the removal of the foreground wedge. To depict the regime where limited frequency resolution and shot-noise become important we show at high $k$ grey points with corresponding error bars that belong to a fiducial $Y=1.00$ higher-resolution simulation. The vertical red dashed line depicts the scale where we roughly expect to enter the non-linear regime. For this non-linear scale we use the relation $k_\mathrm{nl}=0.2\left(1 + z \right)^{2/3}$hMpc$^{-1}$~\citep{Smith:2002dz,Obuljen:2017jiy}. Neglecting scales above this non-linear scale also safely removes the regime where we are dominated by shot-noise at high $k$. For comparison with the power spectra we show the ratio of the corresponding growth rate evolution $f\sigma_8\left(z\right)$ for our two examples of $Y\neq1$ with respect to our fiducial case of $Y=1$, normalised at $z_{cmb}$, in appendix~\ref{app:growth}.

The two cases of stronger and weaker gravitational force $Y=1.01$ and $Y=0.99$, respectively, that deviate from the GR expectation are distinguishable at high signal-to-noise over more than two decades in scales at redshift $z=7$. Here, reionisation is more advanced for stronger gravitational force with $Y=1.01$ and less advanced for gravity weakened with $Y=0.99$ as compared to GR. The different scenarios that are sufficiently close to GR to be in accordance with present-day constraints might also be distinguishable through different large scale behaviour at higher redshift, as for example at $z=10$. A similar picture is found for values of $\alpha$ higher (lower) than one, where reionisation has progressed more (less) as compared to the fiducial case. This underlines the importance of 21cm fluctuations measurements during the EoR in order to constrain modifications of gravity for a wide range of redshifts.

\section{Fisher forecast}\label{sec:Fisher}
\subsection{Cosmological parameters}
In this section we forecast constraints on general modifications of gravity achievable with future 21cm intensity mapping experiments at redshifts of the EoR.
We perform a Fisher matrix forecast for the set of cosmological parameters $\left(\Omega_\mathrm{m,0}, w_0, w_\mathrm{a}, Y, \alpha \right)$. All parameters are assumed to be constant and scale-independent in this first analysis. Note that this does not mean that e.g. $Y$ cannot vary outside of the EoR, as we only constrain our parameters during this epoch. We take parameters for a standard $\Lambda$CDM cosmology with GR as our fiducial model with $\left(\Omega_\mathrm{m,0}, w_0, w_\mathrm{a}, Y, \alpha \right)=\left( 0.308, -1.0,0.0 ,1.0,1.0 \right)$. We investigate both cases of keeping the reionisation history as predicted by astrophysics-related parameters fixed to our fiducial model as stated in eq.~(\ref{eq:iparam}), as well as varying them alongside with cosmological parameters. We also test for different cuts in $k$-space.

For each set of parameters and at each redshift analysed, the corresponding 21cm emission was simulated similar to the examples depicted in section~\ref{sec:sim}, using a modified version of 21cmFAST that includes solving for the growth of linear perturbations as outlined in section~\ref{sec:growth}. 21cm power spectra are then extracted.
We combine constraints for measurements of the 21cm power spectrum in 6 redshift bins from $z=6$ to $z=11$ in steps of $ z=1.0$, accounting for redshift uncertainties of $\Delta z=0.5$. 
For our error estimate we assume a SKA stage 1 like intensity mapping survey to measure the 21cm power spectrum, with survey characteristics as listed in table~\ref{tab:exp}. Cosmic variance, thermal noise and instrumental resolution are accounted for. We  explore as well the case of the 21cm foreground wedge removed, which amounts to assuming perfect foreground removal in the window outside of this wedge; for a description see section~\ref{sec:Pk}. 

For the measurement of 21cm power spectra with an intensity mapping experiment the Fisher matrix is given by
\begin{equation}
F_{ij} = \sum_{z,k} \frac{\Delta k k^2 V_\mathrm{sur}}{4\pi^2} \frac{\partial \tilde{\Delta}_{21}^2(z,k)}{\partial p_{i}} Cov^{-1}(z,k)  \frac{\partial \tilde{\Delta}_{21,l}^2(z,k)}{\partial p_j} ,  
\end{equation}
where $\tilde{\Delta}_{21}$ denotes the dimensionless 21cm power spectrum, derived after parameters $p_i$, and $Cov\left(z,k \right)$ is the covariance matrix for our error modelling as given by eq.~(\ref{eq:sigma21}). For the derivation of the corresponding power spectra we used simulations of box size 300$\,$Mpc, while accounting for the expected angular size of the SKA stage 1 survey of roughly 100$\,$deg$^2$ and the corresponding survey volume by leaving out scales in the sum above that cannot be accessed by the survey. 

\begin{figure}
{\setlength{\tabcolsep}{0.05cm}
\begin{tabular}{cccc}
\includegraphics[width=0.23\columnwidth]{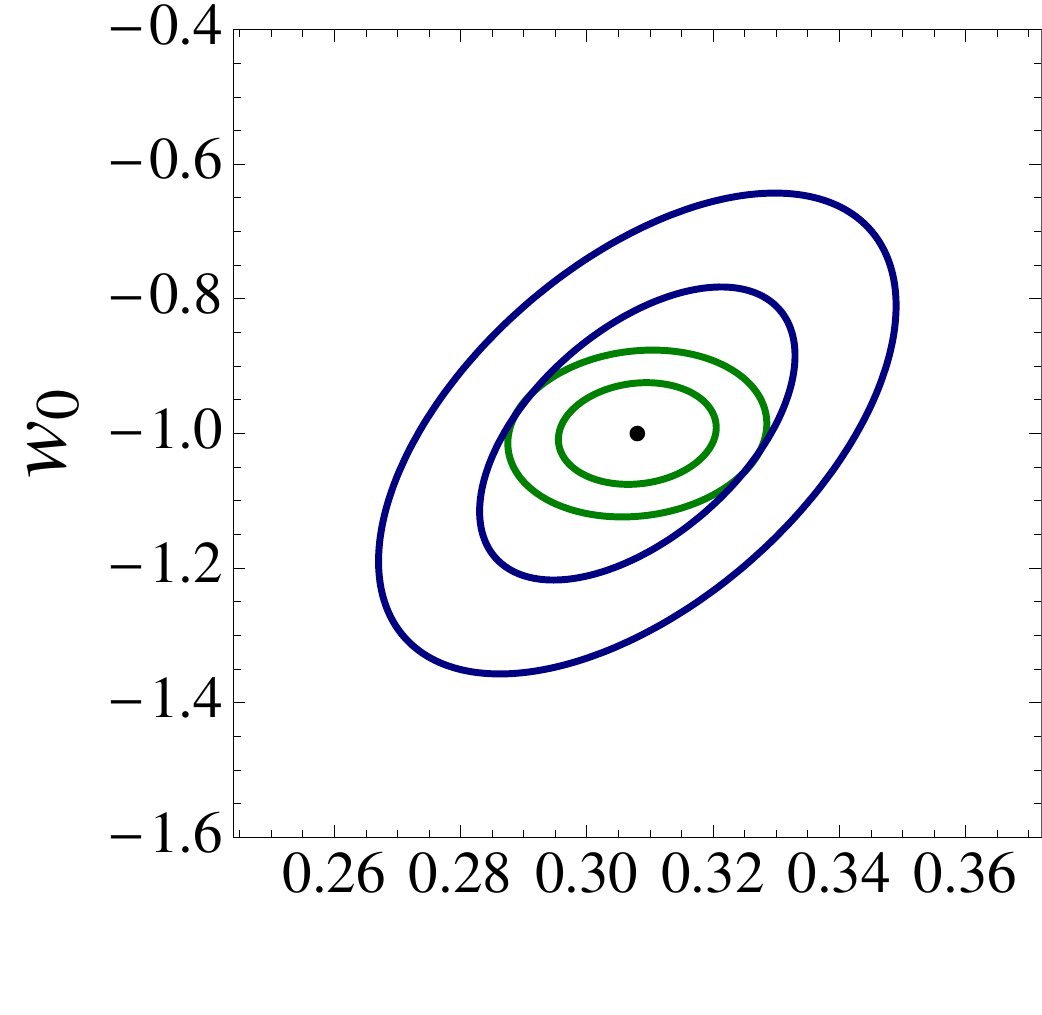}\hfill&\hfill &\hfill &  \\

\includegraphics[width=0.23\columnwidth]{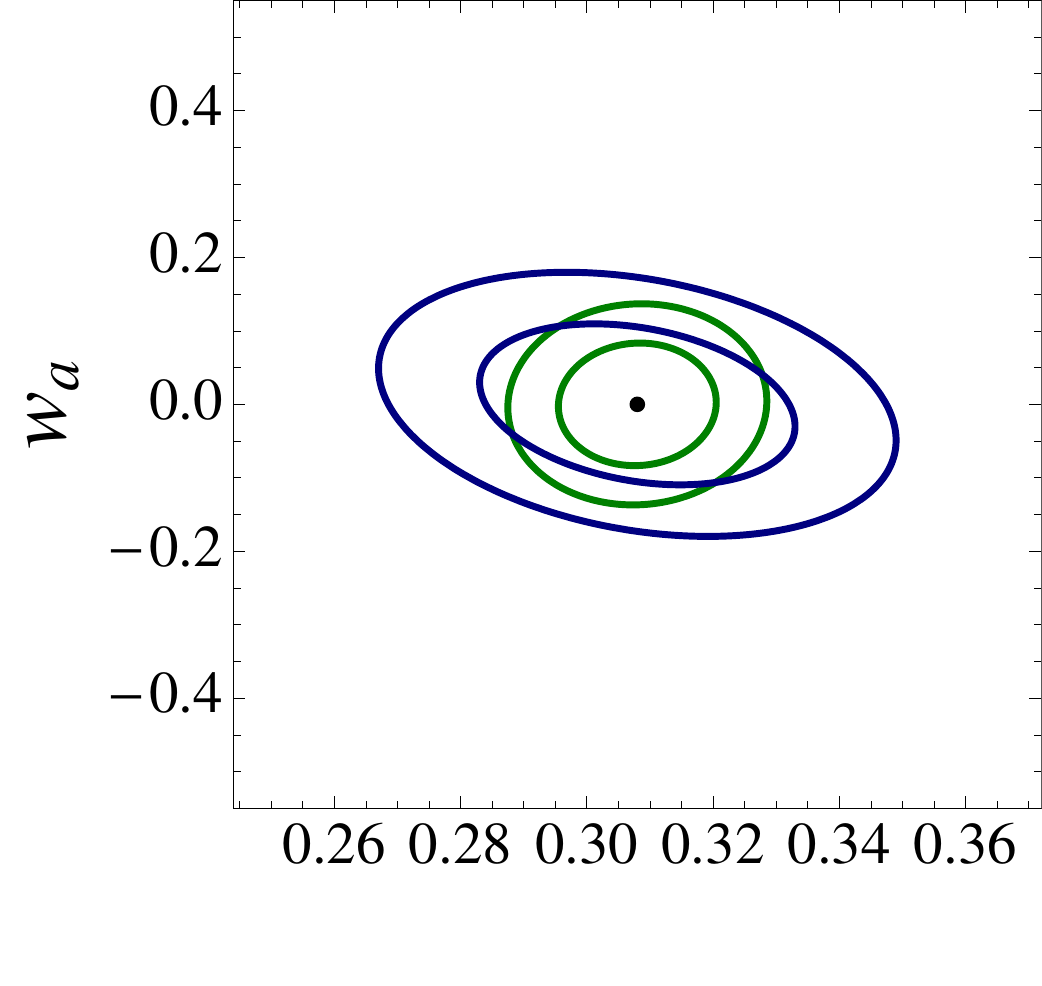} \hfill &
\includegraphics[width=0.23\columnwidth]{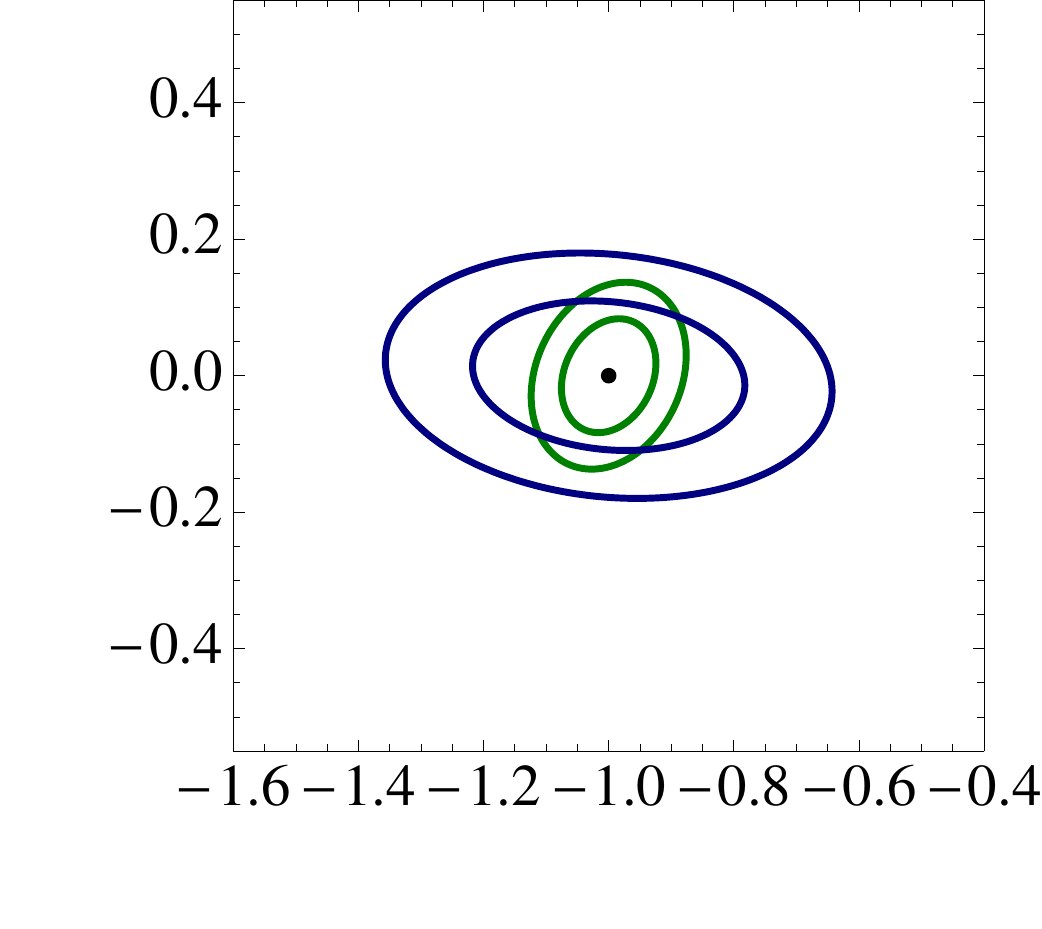}
 \hfill &  \\

\includegraphics[width=0.23\columnwidth]{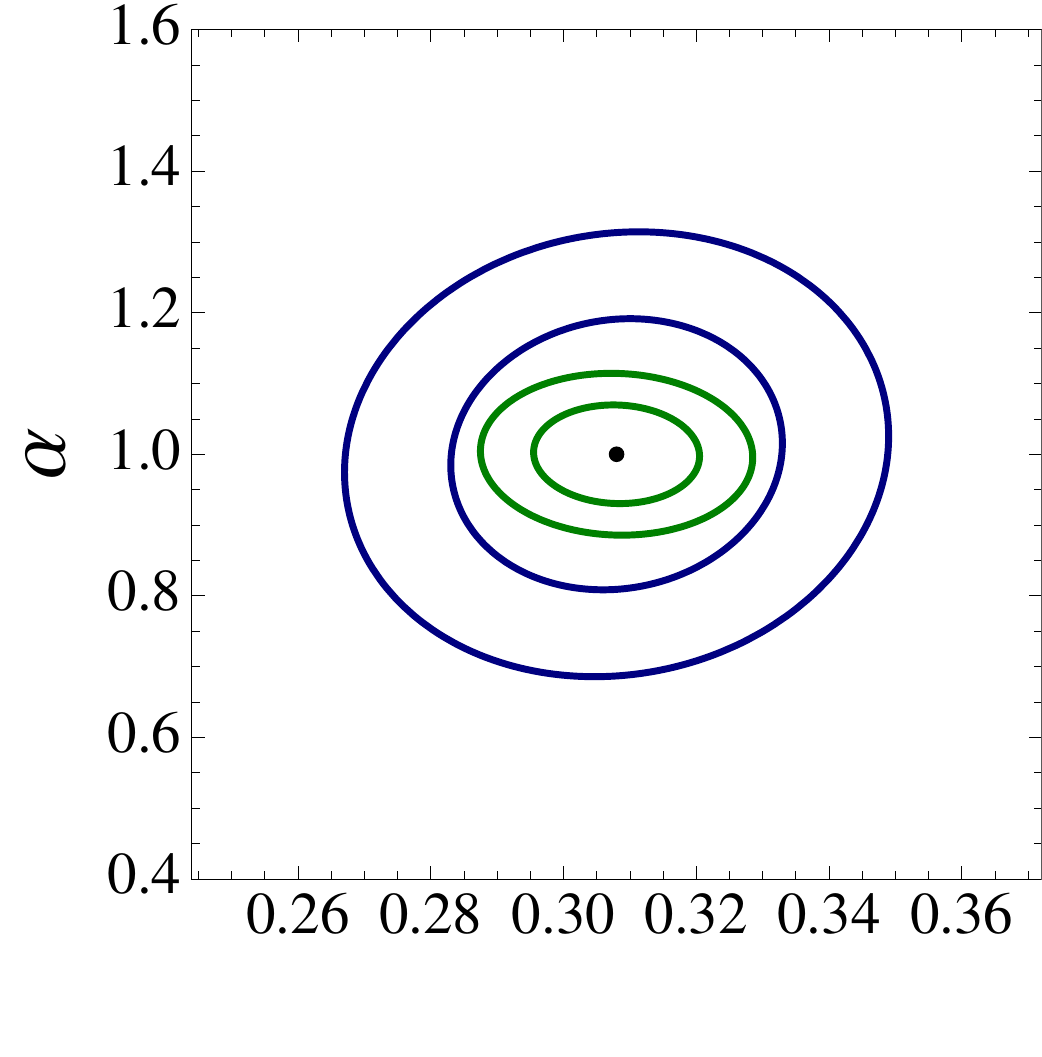} \hfill &
\includegraphics[width=0.23\columnwidth]{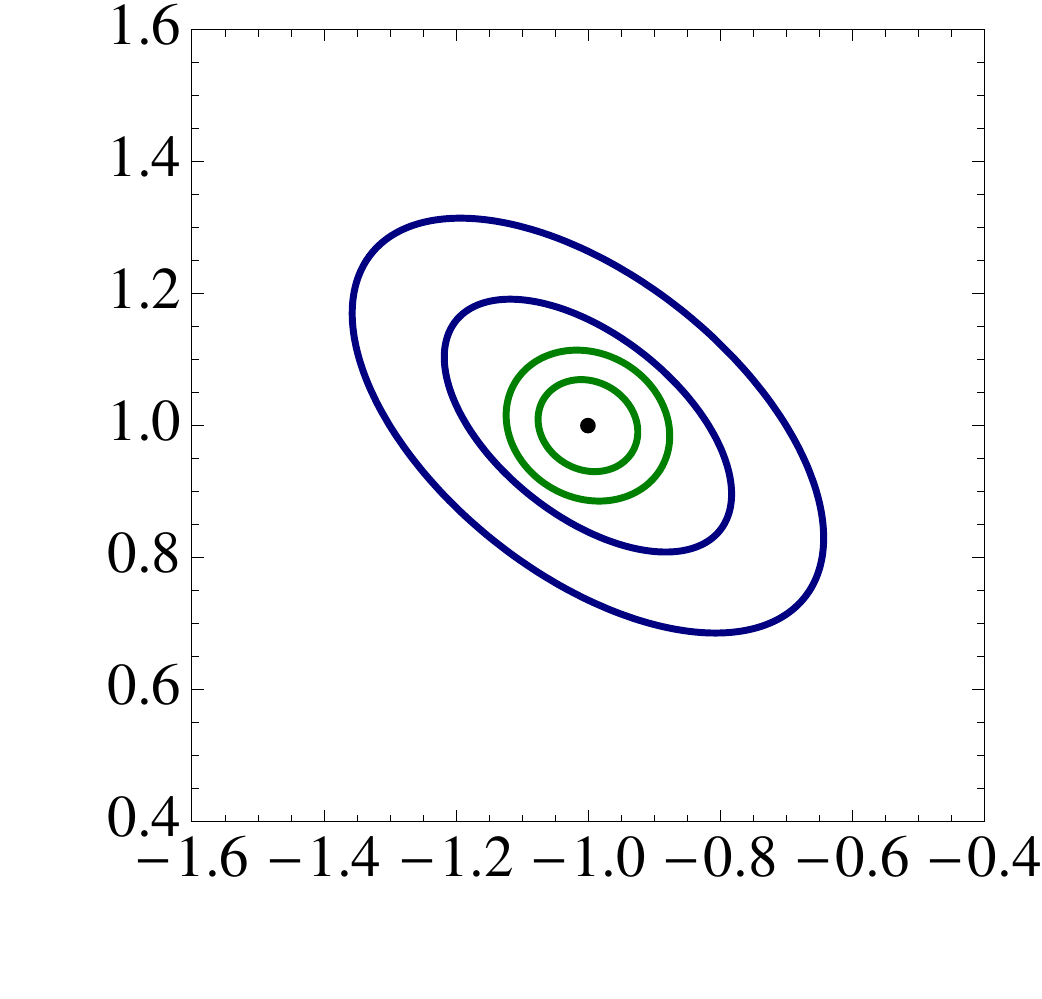}   \hfill&
\includegraphics[width=0.23\columnwidth]{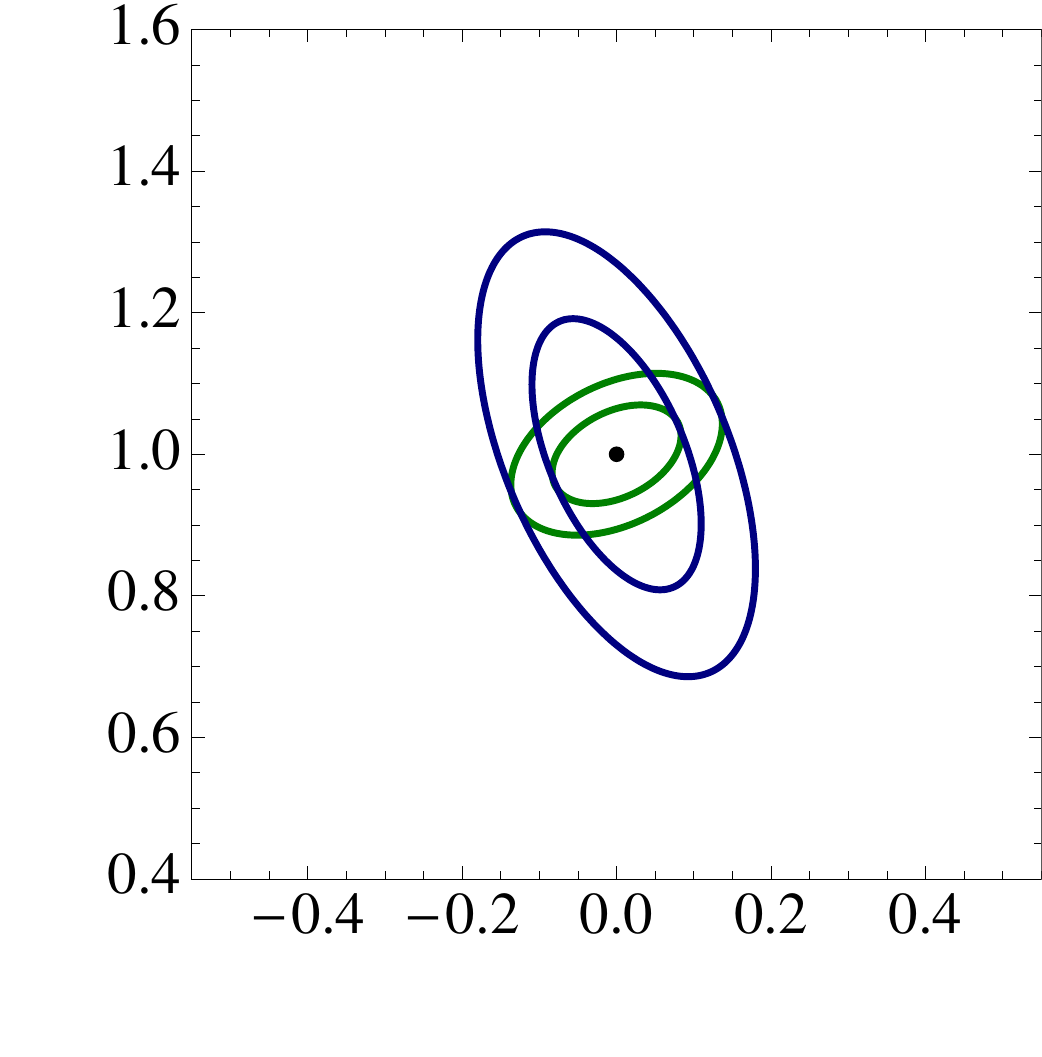} \hfill& \\

\includegraphics[width=0.23\columnwidth]{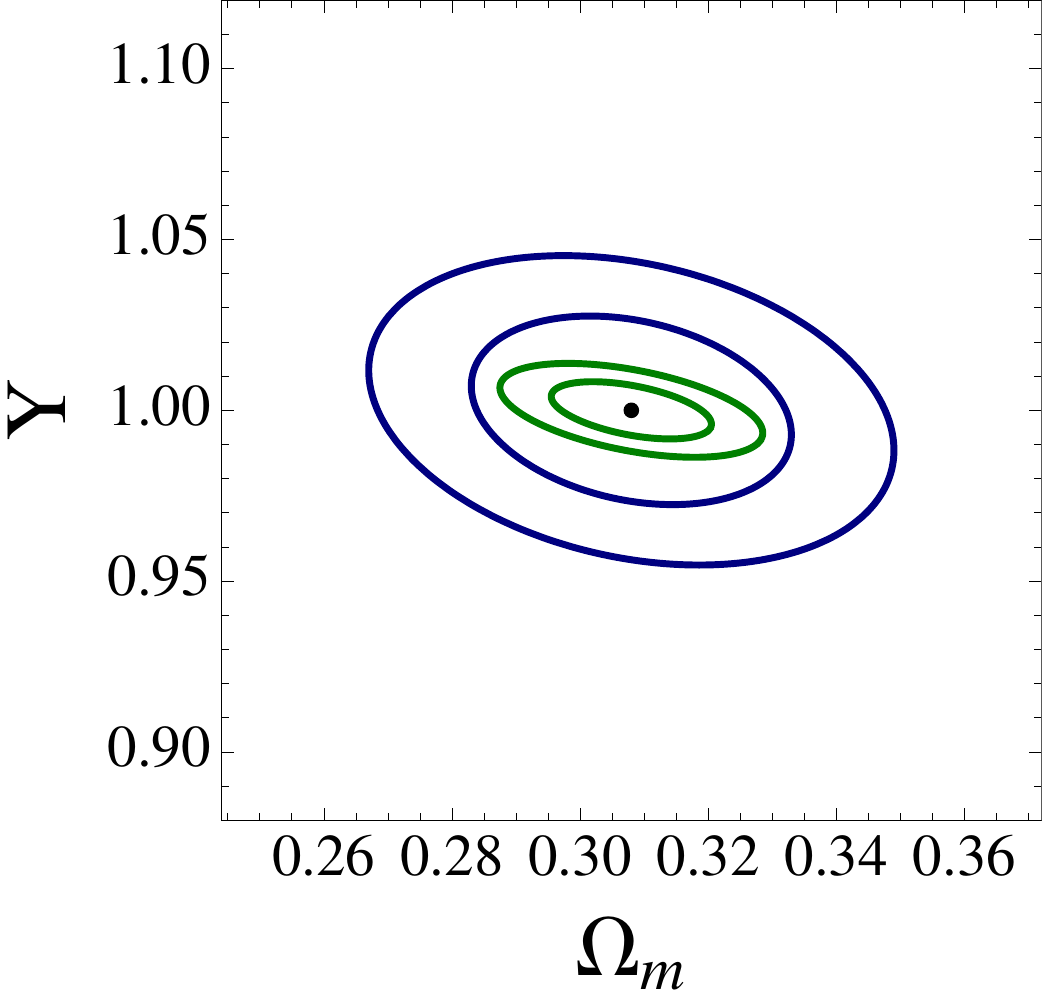}\hfill&
\includegraphics[width=0.23\columnwidth]{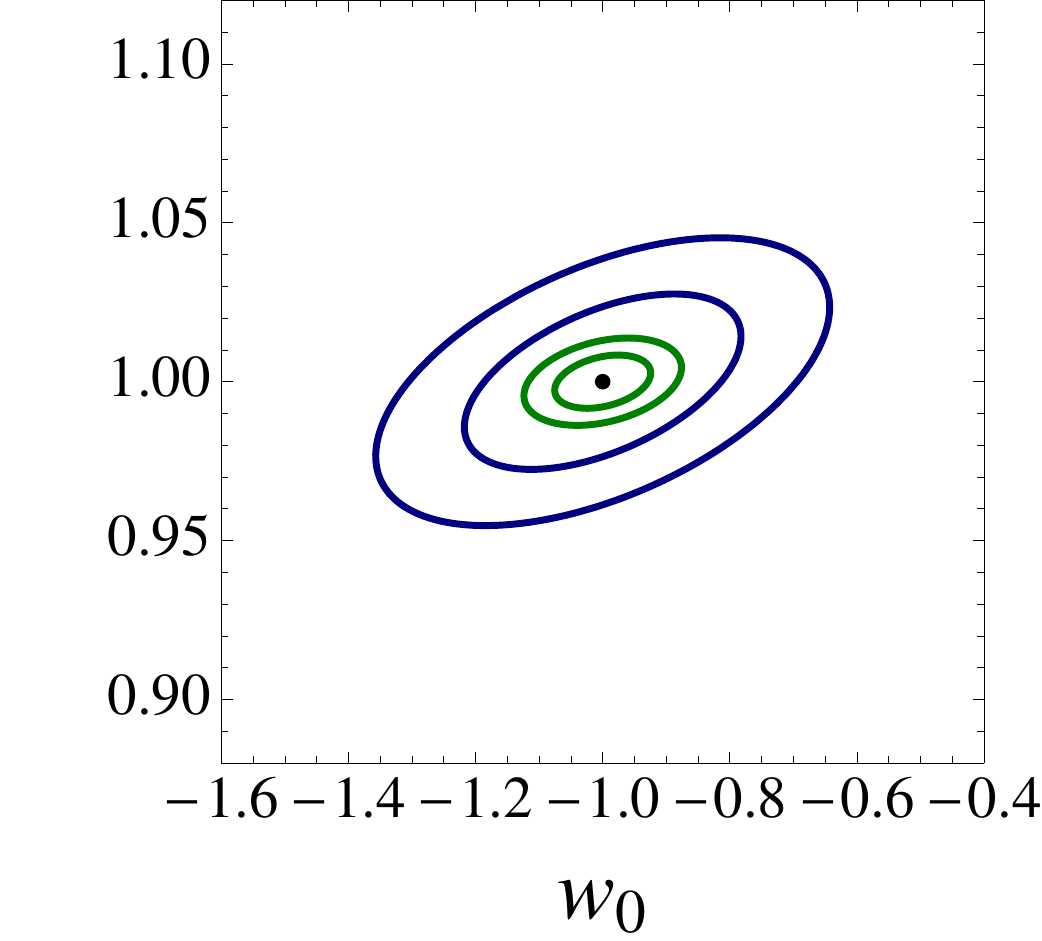} \hfill&
\includegraphics[width=0.23\columnwidth]{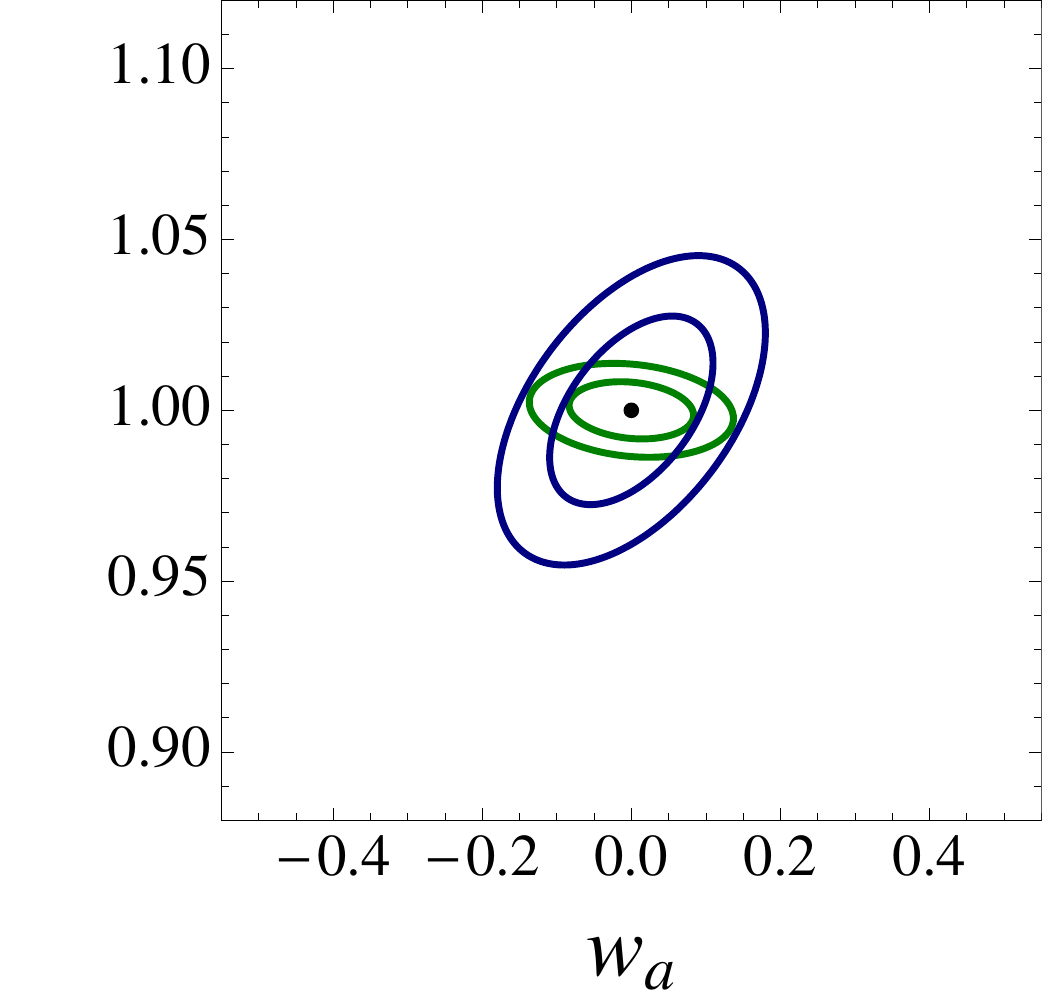} \hfill &
\includegraphics[width=0.23\columnwidth]{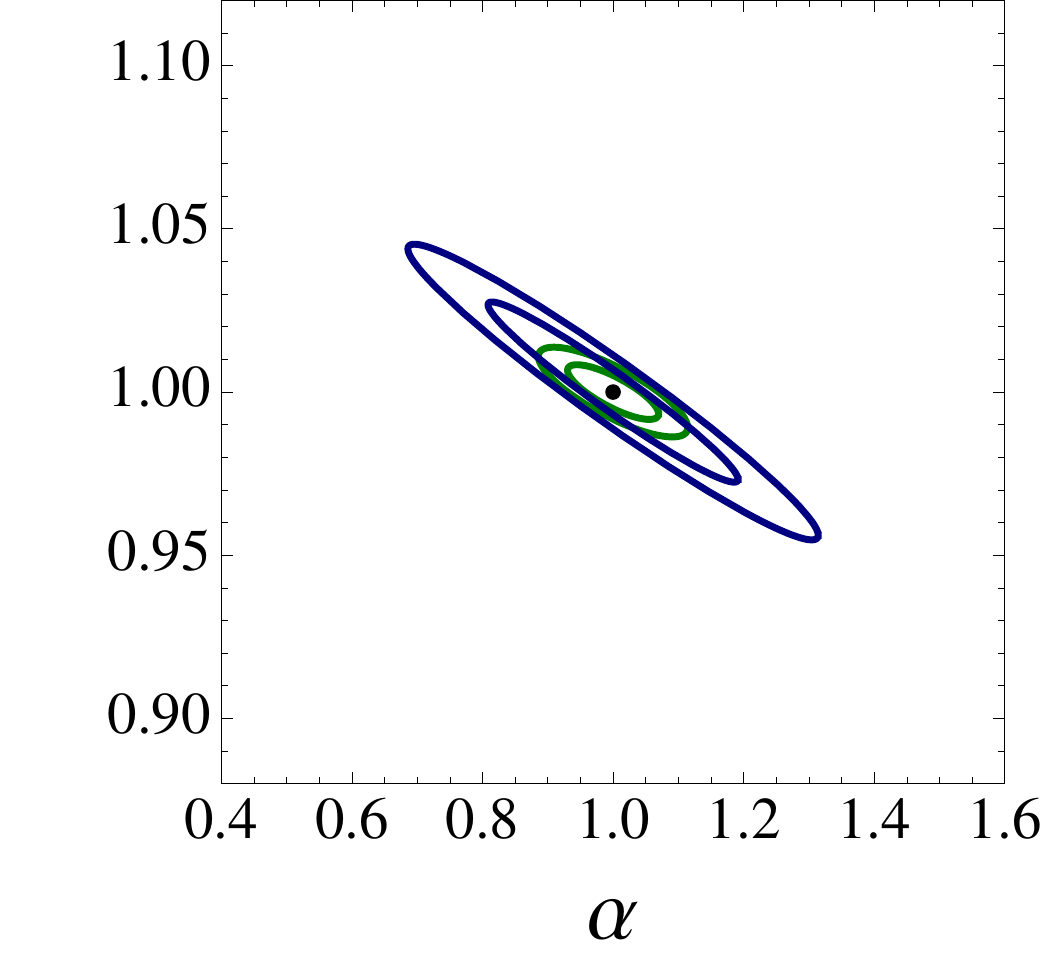}
\end{tabular}
}
\caption{Two-dimensional 1- and 2-$\sigma$ confidence contours, as derived by Fisher forecasting for the set of parameters $\left(\Omega_\mathrm{m,0}, w_0, w_\mathrm{a}, Y, \alpha \right)$. Green contours assume no foregrounds, blue contours assume the removal of the foreground wedge in the analysis. The fiducial model assumes standard $\Lambda$CDM cosmology and GR with $\left(\Omega_\mathrm{m,0}, w_0, w_\mathrm{a}, Y, \alpha \right)=\left( 0.308, -1.0,0.0 ,1.0,1.0 \right)$ (black dots). For more details please see section~\ref{sec:Fisher}.}
\label{FIG:Fisher_CMB}
\end{figure}

For the combination of 21cm power spectra measured in 6 redshift bins from $z=6$ to $z=11$ (note that both the first and the last bin only marginally improve the constraining power on the parameters) and with errors derived as described above to include cosmic variance, thermal noise and instrumental resolution, figure~\ref{FIG:Fisher_CMB} shows the corresponding confidence contours derived for our set of parameters $\left(\Omega_\mathrm{m,0}, w_0, w_\mathrm{a}, Y, \alpha \right)$, both with (in blue) and without (in green) the removal of the 21 cm foreground wedge. As expected, the optimistic case of perfect foreground treatment yields smaller confidence contours as compared to the case of cutting away the foreground wedge, while also impacting some of the parameter correlations.

In table~\ref{tab:Fisher_CMB} the corresponding marginalised 1$\sigma$ confidence limits are shown for the set of 5 parameters that determine our cosmology, both without and with foreground removal, as well as for the parameter set extended to our three reionisation model parameters. 
As we can see here, SKA stage 1 type intensity mapping constraints on standard parameters, here the matter density $\Omega_\mathrm{m,0}$ and the dark energy equation of state $w_0$ parameterised to evolve with redshift via $w_\mathrm{a}$, are competitive for example with upcoming big galaxy surveys at lower redshifts with constraints at the percent level for $\Omega_\mathrm{m,0}$ and $w_0$, respectively. The time evolution of the dark energy equation of state $w_\mathrm{a}$ at redshifts of the EoR will at least be measurable at the ten percent level, as well as general modified gravity parameters $Y$ up to the sub-percent and $\alpha$ at the percent to ten percent level, respectively, in our optimistic scenario. Tomographic information due to the large redshift coverage from $z=6-11$ adds to the scale-dependent information encoded in the 21cm power spectra (indicated by the sensitivity to cuts in k-space). This for example conspires to measure the parameters for an evolving dark energy equation of state individually with a precision comparable to large galaxy surveys (that cover a smaller redshift range). While small deviations from $Y=1$ that point to modified gravity become detectable at high redshifts, also deviations from $\alpha=1$ can point to Brans-Dicke or certain coupled dark energy type of models. In the Brans-Dicke case the most stringent constraints come from solar system tests, but they only constrain $z=0$ locally and assume no (effective) screening. For some coupled dark energy models a detection of $\alpha>1$ would reveal a non-negligible early dark energy density.

 \renewcommand{\arraystretch}{1.2}
\setlength{\tabcolsep}{7pt}
    \begin{table}
    \centering
    \begin{tabular}{|c| c| c| c| c| c| c| c| c|}
    \hline 
  &   $\Omega_\mathrm{m,0} $   & $w_0$  & $w_\mathrm{a}$   & $ Y$  &  $\alpha$  & $R_\mathrm{mfp}$ [Mpc] & $T_\mathrm{vir}$ [K] & $\zeta$ \\
    \hline
    fiducial &  0.308  &  -1.0 &  0.0 &  1.0 &  1.0 & 20 &  $3\times10^4$ & 20 \\ 
    \hline
1$\sigma$ error &  0.008  &  0.05 &  0.06 &  0.006 &  0.06 &  - & - & - \\ 
  \hline
  1$\sigma$ error +'w' &  0.016  & 0.144 & 0.072  & 0.018  & 0.126 &  - & - & - \\ 
  \hline
  1$\sigma$ error &  0.022  &  0.023 &  0.024 &  0.013 &  0.119 &  2.0 & 31.7 & 2.8 \\ 
  \hline
      1$\sigma$ error +'w' &  0.015  & 0.202 & 0.297  & 0.016 & 0.103 & 1.3 & 31.8 & 2.6  \\ 
  \hline
      1$\sigma$ error + $k_\mathrm{shot}$ &  0.075  & 0.260 & 0.443  & 0.039  & 0.341 &  - & - & - \\ 
  \hline
      1$\sigma$ error + $k_\mathrm{lin}$ &  0.307  & 0.841 & 0.548  & 0.116  & 0.879 &  - & - & - \\ 
  \hline
         \end{tabular}
          \caption{Marginalised 68.3 percent confidence intervals for an intensity mapping experiment type SKA stage 1 for 6 redshift bins $z=6$ to $z=11$ in steps of one in redshift to account for an expected redshift uncertainty of $\Delta z=0.5$. Shown are values assuming perfect foreground removal, as well as values for the so-called foreground wedge removed denoted by +'w', and cuts above the shot-noise scale and above linear scales, denoted by +$k_\mathrm{shot}$ and +$k_\mathrm{lin}$. For more details on the Fisher matrix forecast see section~\ref{sec:Fisher}. }
    \label{tab:Fisher_CMB}
    \end{table}

An important point we would like to make with semi-numerical simulations of the 21cm emission signal in this paper, is that high-redshift intensity mapping probes can give unique constraints on general modifications of gravity at redshifts of reionisation. Including also the reionisation model parameters in our analysis yields still comparable constraints on cosmological parameters weakened by a factor of two to four, while simultaneously constraining the typical halo virial temperature $T_\mathrm{vir}$ at the percent, and the mean free path $R_\mathrm{mfp}$ as well as ionising efficiency $\zeta$ at the ten percent level. Combining tomographic bins of 21cm intensity mapping is thus efficient in breaking parameter degeneracies, with most constraining power coming from redshifts $z=7-10$ when the power spectrum shape is evolving fastest.

We note that errors with wedge removed, denoted by +'w' in table~\ref{tab:Fisher_CMB}, are always larger, which is expected, except when the parameter set is extended to reionisation parameters, for the $\Omega_\mathrm{m,0}$, $\alpha$, $R_\mathrm{mfp}$ and $\zeta$ parameters. The reason being that correlations between parameters turn out to be far lower in the case when the wedge is removed, translating to slightly smaller errors for some parameters.
Also, including the three ionisation parameters, the errors always increase, except for $Y$ with wedge removed.
Here the $Y$ parameter for the wedge removed is more correlated with $w_\mathrm{a}$ for cosmological parameters-only, and an anti-correlation with $R_\mathrm{mfp}$ arises. As correlations between parameters seem to play a crucial role for the constraints, we shortly examine these in the following section.

We also note, that for more conservative cuts regarding the linear nature of our modelling, denoted by +$k_\mathrm{shot}$ and +$k_\mathrm{lin}$, a cut above the shot-noise scale of roughly 1$\,$Mpc$^{-1}$ and a cut that estimates the validity of linear perturbations in $\Lambda$CDM as in the previous section, respectively, constraints are weakened considerably, see again table~\ref{tab:Fisher_CMB}. For the cut of shot-noise dominated scales, constraints are weakened by a factor of five to ten, while for linear scales-only cuts, the constraints one some parameters can be weakened by an additional factor of three to four. Neglecting (mildly) non-linear has a larger impact on constraints than foreground cuts or the addition of reionisation parameters. It will therefore be crucial in the future to include this modelling  for models beyond GR.

\begin{figure}
\begin{center}
\includegraphics[width=0.45\columnwidth]{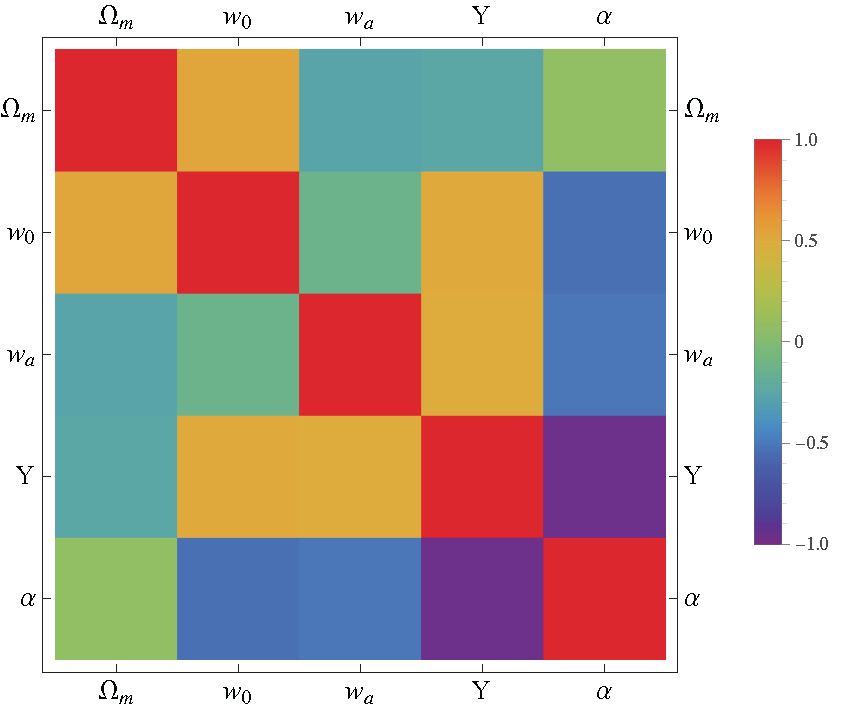}
\includegraphics[width=0.45\columnwidth]{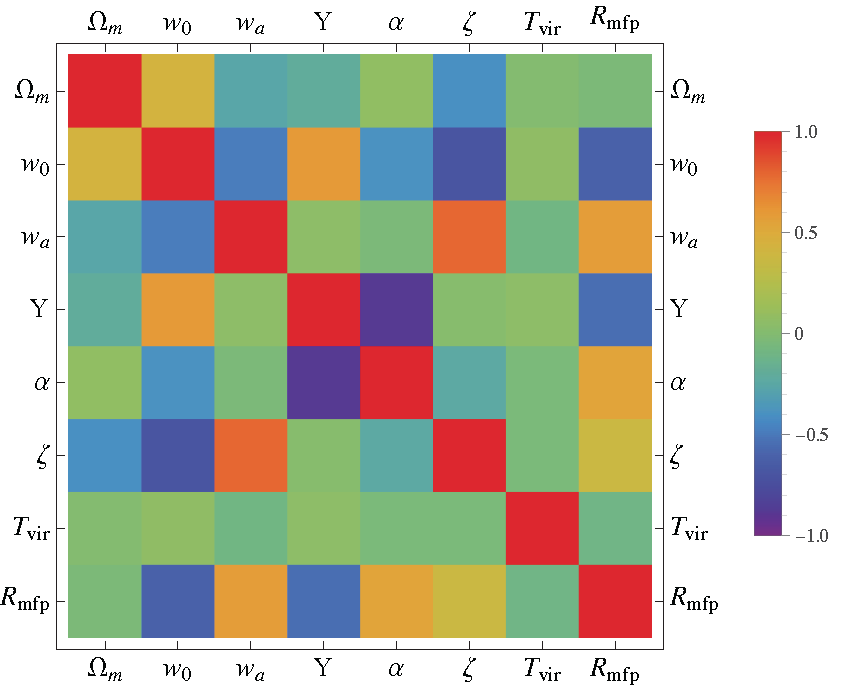}
\caption{Correlation matrices for our Fisher analysis. Left: Cosmological parameters only. Right: Both cosmological and reionisation model parameters. The foreground wedge was removed in this analysis.}
\label{FIG:corrR}
\end{center}
\end{figure}

\subsection{Correlation with reionisation parameters}
To check how strongly the assumption of keeping the reionisation model parameters fixed will impact the Fisher forecast of our cosmology, besides slightly widening the 1$\sigma$ marginalised errors as presented in the previous section, we calculated the covariance matrix ${\bf C}$ as the inverse of the Fisher matrix for the full parameter vector $p=\left( \Omega_\mathrm{m,0}, w_0, w_\mathrm{a}, Y, \alpha, \zeta, T_\mathrm{vir}, R_\mathrm{mfp}\right)$ that also includes reionisation parameters, for comparison with the cosmological parameters-only case. The correlation matrix ${\bf P}$ is then calculated as 
\begin{equation}
P_{ij} = \frac{C_{ij}}{\sqrt{C_{ii} C_{jj}}} .
\end{equation}
This correlation matrix $P$ is a measure for the presence of non-diagonal elements in the covariance matrix and is 1 for perfect correlation and -1 for perfect anti-correlation between two elements of our parameter vector p. In figure~\ref{FIG:corrR} we show the correlation matrices for the full parameter vector of cosmological and reionisation parameters (right panel) as well as cosmology-only (left panel), where in both cases we included foreground removal in the analysis. We see that correlations between cosmological parameters are only slightly affected by adding reionisation model parameters in the analysis. The virial temperature turns out to be uncorrelated with cosmological parameters, while both ionising efficiency and mean free path show correlation and anti-correlation with some of the cosmological parameters. 

As one example, the modified gravity parameter $Y$ that regulates the strength of gravity as compared to GR is anti-correlated with the mean free path, with stronger gravity leading to a somewhat shorter mean free path, but shows little correlation with both the typical halo virial temperature and the ionising efficiency. Similarly, $w_0$ is anti-correlated with $\zeta$ and $R_\mathrm{mfp}$, intuitively translating to a situation where, for less accelerated expansion, a smaller mean free path or smaller ionising efficiency are required to yield the same reionisation history. Measurements of reionisation parameters, for example by including information from other lines than 21cm and cross-correlations~\citep{2017ApJ...848...52H}, can improve the constraints on cosmology, but this effect is at the moment smaller as compared to the impact of including or excluding non-linear scales. As for galaxy surveys~\citep{Casas:2017eob}, the modelling of non-linear scales might be important for de-correlating parameters.

\section{Conclusion}
In this work we have included a scenario of general modifications to gravity into semi-numerical simulations of 21cm emission during the EoR. We did so by modifying the growth history, adding modified gravity parameters $Y$, for deviations from a standard Poisson equation, and $\alpha$, for initial conditions that deviate from early matter domination, together with a time-varying dark energy equation of state, on top of the standard $\Lambda$CDM and GR scenario. 
The corresponding 21cm power spectra were derived from simulation boxes of cosmological volumes. 

We showed with a Fisher matrix forecast, that the constraints on modified gravity parameters, alongside with the standard matter density $\Omega_\mathrm{m,0}$ and dark energy equation of state parameters $w_0$ and $w_\mathrm{a}$, can reach the sub-percent level for example for a deviation of $Y$ from a GR Poisson equation, when assuming an intensity mapping experiment of the type of SKA stage 1. When disregarding scales beyond the shot-noise scale, reaching for example 4$\%$ constraints on $Y$ is within reach. Whether these levels can be reached, crucially depends on the foreground treatment and modelling of the (mildly) non-linear regime. 

Extending the set of parameters from cosmology-only to reionisation model parameters for our fiducial model, chosen to match available present-day constraints both for cosmology and astrophysics, does not degrade parameter constraints significantly. Here we note, that having different tomographic bins in redshift is a deciding factor for breaking parameter degeneracies, while most constraining power comes from redshifts 7 to 10 when the power spectrum shape is changing fastest. We also stress here that modelling the (mildly) non-linear regime turns out to be more crucial for stringent constraints on general modifications of gravity, than for example degeneracies with astrophysical model parameters.

To conclude, competitive constraints on modifications of gravity at redshifts during the EoR untested so far are possible for the first time with tomographic 21cm experiments. With the 21cm signal being more sensitive to deviations from GR than growth alone, as well as reaching constraints competitive with for example galaxy clustering surveys at low redshifts, even testing time- or scale-dependence for modifications of gravity at high redshift comes within reach of upcoming intensity mapping experiments.

 \acknowledgments
We thank the DFG for supporting this work through the SFB-Transregio TR33 ``The Dark Universe". We thank the referee for useful comments and suggestions. C.H. thanks Matteo Viel for useful discussions.

\appendix
\section{Extreme examples for modified gravity during reionisation}\label{sec:extreme}
\begin{figure}
\begin{center}
\includegraphics[width=0.32\columnwidth]{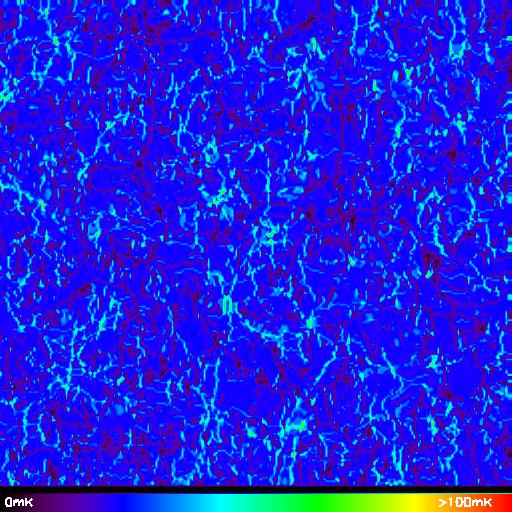}
\includegraphics[width=0.32\columnwidth]{sBox_21cm_z010_Y1.jpg}
\includegraphics[width=0.32\columnwidth]{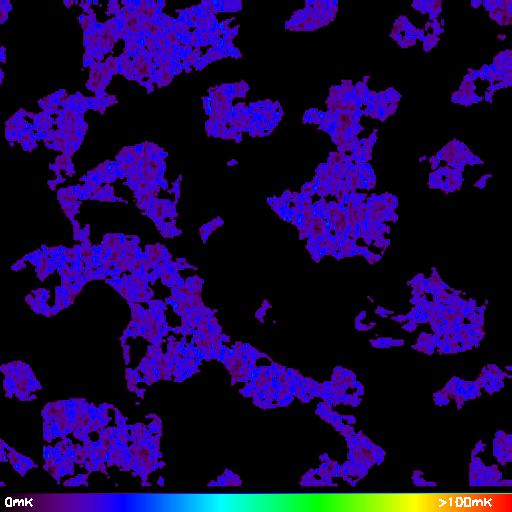}
\caption{Examples of 300$\,$Mpc simulation boxes of 21cm emission at redshift $z=10$, for $w_0=-1$, $w_\mathrm{a}=0$, $\Omega_\mathrm{m,0}=0.308$, $\alpha=1$ and varying $Y=2.0$ (left), $Y=1.0$ (middle) and $Y=0.8$ (right); here the growth function was normalised to the same value at $z=0$ for all models, and therefore all have the same rms mass fluctuations $\sigma_8=0.815$, i.e., fixed to the fiducial value, at 8/h$\,$Mpc scales at $z=0$.}
\label{FIG:Box21_Y_z010}
\end{center}
\end{figure}

Here simulation boxes of 21cm emission are depicted for extreme values of the parameter $Y$ that enters the Poisson equation. The goal is not to show scenarios in accordance with present-day observational constraints, but to show clearly how modified gravity affects the reionisation history and therefore the signal in 21cm emission by altering the growth of structures. 
 Figure~\ref{FIG:Box21_Y_z010} shows simulation boxes, 300$\,$Mpc box size, of fluctuations in 21cm brightness temperature at redshift $z=10$ for extreme values of $Y=2.0$ (left) and $Y=0.8$ (right) alongside with our fiducial model $Y=1.0$ (middle panel) for $w_0=-1$, $w_\mathrm{a}=0$, $\Omega_\mathrm{m,0}=0.308$ and $\alpha=1$ fixed. With neutral fractions of approximately 30$\%$, 87$\%$ and 100$\%$ from left to right, already `by eye' the effect of a weakened versus strengthened gravitational force becomes obvious. We can see, that depending on the value of $Y$, reionisation has either not started yet ($Y=2.0$), has begun with a fraction of $87\%$ of hydrogen still neutral ($Y=1.0$), or has significantly progressed already at $z=10$ with only $30\%$ of the medium still neutral ($Y=0.8$).

\section{Examples for the growth rate evolution f$\sigma_8\left(z\right)$}\label{app:growth}
\begin{figure}
\begin{center}
\includegraphics[width=0.7\columnwidth]{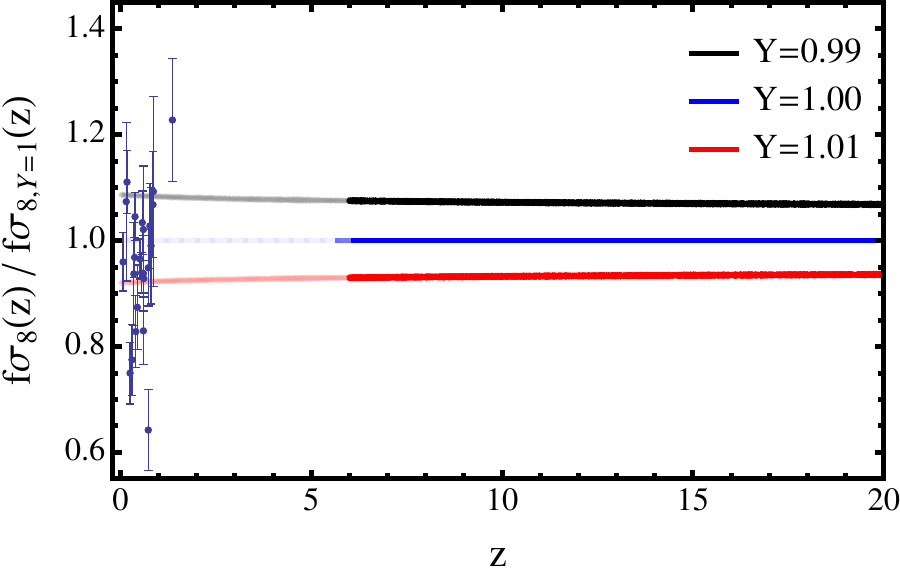}
\caption{Fractional difference for growth rate evolution f$\sigma_8\left(z\right)$ with respect to GR+$\Lambda$CDM ($Y=1$) for $Y=0.99$ and $Y=1.01$, for our simulation examples from section~\ref{sec:sim}. The growth is evolved via eq.~(\ref{eq:pert2}) and normalised to the same growth at the redshift of recombination. For more details see section~\ref{sec:growth}. Despite our measurements not being sensitive to low redshifts, we show f$\sigma_8\left(z\right)$ down to $z=0$ together with f$\sigma_8$ data points from RSD measurements~\cite{Kazantzidis:2018rnb} for comparison.}
\label{FIG:growth}
\end{center}
\end{figure}

We illustrate here the size of the difference in growth rate f$\sigma_8\left(z\right)$ for our exemplary models from section~\ref{sec:sim}, where the normalisation of the growth factor $G(z)$ was fixed at the redshift of recombination. Figure~\ref{FIG:growth} shows the fractional difference with respect to GR+$\Lambda$CDM ($Y=1$) of the growth rate evolution f$\sigma_8\left(z\right)$ during reionisation for the two modified gravity values of $Y=0.99$ and $Y=1.01$ (cosmological parameters other than $Y$ match the fiducial model from section~\ref{sec:growth}). By the time of reionisation around redshifts $z=6-10$, the difference in the growth factor has reached about $2-3\%$ for our models, whereas for f$\sigma_8$ plotted here a difference of $1\%$ in $Y$ translates to a $7-8\%$ difference. For comparison we show the evolution down to $z=0$ together with data points compiled from RSD measurements~\cite{Pinho:2018unz}, where both models are still allowed within the current precision. We stress that our 21cm measurements are sensitive to the integrated effect down from the CMB redshift, but not to post-reionisation redshifts.
In comparison to the difference in corresponding 21cm power spectra in section~\ref{sec:Pk}, we note that the 21cm signal is more sensitive to deviations from GR than growth alone. 

\bibliographystyle{JHEP}
\bibliography{ref_growth.bib}

%
%
%
%
%
%
%
%
\end{document}